\documentclass[amsmath,amssymb,pre,twocolumn,floatfix]{revtex4-1}

\usepackage{color}
\usepackage{graphicx}
\usepackage{dcolumn}
\usepackage{times}
\usepackage{soul}
\usepackage{siunitx}
\usepackage{afterpage}
\usepackage{xcolor}
\usepackage{tabularx}
\usepackage{placeins}
\usepackage[breaklinks=true,colorlinks,citecolor=blue,linkcolor=blue,urlcolor=blue]{hyperref}
\newcolumntype{b}{>{\hsize=.12\textwidth}X}
\newcolumntype{s}{>{\hsize=.15\textwidth}X}

\begin{document}
	
	\title{ Mechanochemical induction of wrinkling morphogenesis on elastic shells}
	\author{Andrei Zakharov}\email{azakharov@ucmerced.edu}
	\author{Kinjal Dasbiswas}\email{kdasbiswas@ucmerced.edu}
	\affiliation{Department of Physics, University of California, Merced, CA 95343, USA}
	
	\begin{abstract}
		Morphogenetic dynamics of tissue sheets require coordinated cell shape changes regulated by global patterning of mechanical forces. Inspired by such biological phenomena, we propose a minimal mechanochemical model based on the notion that cell shape  changes are induced by diffusible biomolecules that influence tissue contractility in a concentration-dependent manner -- and whose concentration is in turn  affected by the macroscopic tissue shape. We perform computational simulations of thin shell elastic dynamics to reveal propagating chemical and three-dimensional deformation patterns arising due to a sequence of buckling instabilities.  Depending on the concentration threshold that actuates cell shape change, we find qualitatively different patterns. The mechanochemically coupled patterning dynamics are distinct from those driven by purely mechanical or purely chemical factors. Using numerical simulations and theoretical arguments, we analyze the elastic instabilities that result from our model and provide simple scaling laws to identify wrinkling morphologies.
		
		
	\end{abstract}
	
	\maketitle
	
	\section{Introduction}
	
	Morphogenetic events during embryo development involve tissue shape changes that are driven by mechanical forces \cite{thompson_1992, forgacs_2005}. A prototypical example is the folding of sheets of epithelial cells through the constriction and shape change of individual cells by actively contractile forces generated by myosin molecular motors in the cytoskeletal network on the cell surface \cite{lecuit_07, gilmour_17}. The myosin motor activity is in turn triggered by complex chemical signaling cascades secreted by the cells themselves which create a spatiotemporal pattern of active mechanical forces in the tissue sheets  \cite{goldstein_10}. The chemical signaling itself can be affected by mechanical cues such as as forces \cite{labousse} and deformations \cite{farge_11, fernandez_15}. Inspired by the natural spatiotemporal control of shape change in biological tissue sheets, we are motivated to ask how such bio-inspired feedback can be used to realize self-actuated patterning of soft materials. 
	
	Thin elastic plates and shells constitute a fundamental class of soft matter that exhibit sensitive response to stimulii changes because of the geometric nonlinearity of their mechanical properties \cite{Holmes2019}.  Tissue morphogenesis can inspire the design of such slender structures that change shape in a programmable way. Desired shapes can be attained by pre--patterning the internal structure, for example via metric change \cite{kim2012designing,aharoni2018universal,klein2007shaping}, controlled intrinsic curvature \cite{gladman2016biomimetic, zakharov2019programmable}, pre--stress \cite{xie2010tunable}, or be self-organized, 
    for example, through a propagating chemical reaction coupled with mechanical deformations \cite{okuda2018combining,mercker2013mechanochemical,mietke2019self,yoshida2010self}. The latter approach provides reconfigurable conformations in contrast with frozen-in patterns, and resembles the shape changes during morphogenesis with anisotropic deformations including different types of mechanical instabilities, such as creases, wrinkles, folds, and ridges \cite{Li2012}. 
    
    Instabilities in spherical shells are of special interest because of their diverse applications and non-trivial behavior from a mathematical perspective. Examples of thin shells that exhibit wrinkling morphology abound in nature and range from pollen grains and viral capsids to organoids and organs such as brain and developing embryo at the ``blastula'' stage \cite{forgacs_2005} (monolayer of cells arranged in a shell surrounding a fluid-filled lumen). Several studies have examined buckling behaviour of spherical shells due to external pressure \cite{hutchinson2016buckling,paulose_13}, confinement \cite{trushko2020buckling},  differential growth \cite{li2011surface,hohn2015dynamics}, and demonstrated that patterns arising on curved surfaces are quite different from those on planar surfaces. Curvature-controlled structure formation encompasses a broad class of phenomena in physical \cite{stoop2015curvature,cao2008self} , biological \cite{richman1975mechanical} and chemical systems\cite{yin2014surface}.
	
	In comparison to synthetic materials, living matter can show more complex cycles of feedback and control where chemical and mechanical signalling are tightly coupled. Inspired by this inherently mechanochemical basis of pattern formation in biological tissue \cite{howard_11}, we consider the dynamics and steady state of shapes induced by chemical gradients in thin elastic shells, where the chemical is itself affected by the sheet curvature. In analogy with tissue patterning by gradients of morphogens, diffusible biomolecules that induce cell fate changes slowly in a concentration-dependent manner during embryogenesis \cite{wolpert_69}, we have posited ``mechanogens'' as biochemical agents that affect the cell mechanical state \cite{dasbiswas_16, dasbiswas_18}. They can do so by enhancing or relaxing the cell cytoskeletal contractility, which is a more physical change in comparison to the usual genetic changes wrought by morphogen gradients.  Candidate mechanogens could be chemical factors, such as Ca$^{2+}$ ions, proteins, ATP or drugs, that regulate actomyosin contractility and cytoskeletal remodeling. While disentangling the various chemical signals and their interactions with mechanics is challenging \emph{in vivo}, recent experiments \emph{in vitro} with reconstituted cytoskeletal gels that exhibit molecular motor activity-driven buckling and wrinkling \cite{ideses_18, senoussi_19, strubing_20} realize such active elastic processes in a controlled setting. These biological materials, along with synthetic materials such as liquid crystal elastomers \cite{white_15} and gels \cite{ionov2014hydrogel}, in principle, allow for the spatial control of mechanical deformation and shape actuation through external chemical, electrical or optical stimuli \cite{bruegmann2010optogenetic}.
	
	In this work, we explore how in-plane stresses induced in a thin elastic sheet by gradients of chemical signals (``mechanogens'') can be relaxed by energetically less costly out-of-plane deformations. We thus seek to demonstrate how a short range chemical activation leads to long range elastic response and subsequent pattern formation. Unlike models for tissue folding that are based on differential apical-basal constrictions that lead to wedge-shape of the constituent cells \cite{hannezo_14,brinkmann2018post}, our model is based on a spontaneous curvature arising due to in-plane incompatibility between domains of different tension -- an effect that is similar to differential growth and also occurs in principle in contractile biological tissue.  We show that depending on key parameters, such as a threshold of activation of cell mechanical response by the chemical gradient and the thickness of the elastic sheet, we can access qualitatively different patterns such as  ridges and spots. Further, the mechanical feedback on the chemical gradient results in pattern propagation from an initial local region of activation.  While inspired by tissue morphogenesis, these results may also be applicable to synthetic gels provided such a feedback can be set up.  
	
	The structure of the paper is organized as follows. We first define governing equations for a model involving mechanochemical interactions and give an estimation of the time scales that allow simplifying assumptions. In Sec.\ref{Section_Res}, we demonstrate patterns arising due to feedback between chemical production and elastic instabilities, and explore the parameter space identifying qualitatively different stationary shapes. In Sec.\ref{Section_Buckling}, we discuss the buckling and wrinkling instabilities, and provide simple scaling laws for the observed patterning. In the Appendix, we describe the discretized elastic energy, the numerical methods applied to perform simulations of model dynamics, and provide supplementary simulation results obtained for parameter values different from those in main text.
	
	\section{Mechanochemical model for tissue shape change}\label{Section_Model}
	
\begin{figure}[t]
		\centering
		(a)\\
		\includegraphics[width=0.42\textwidth]{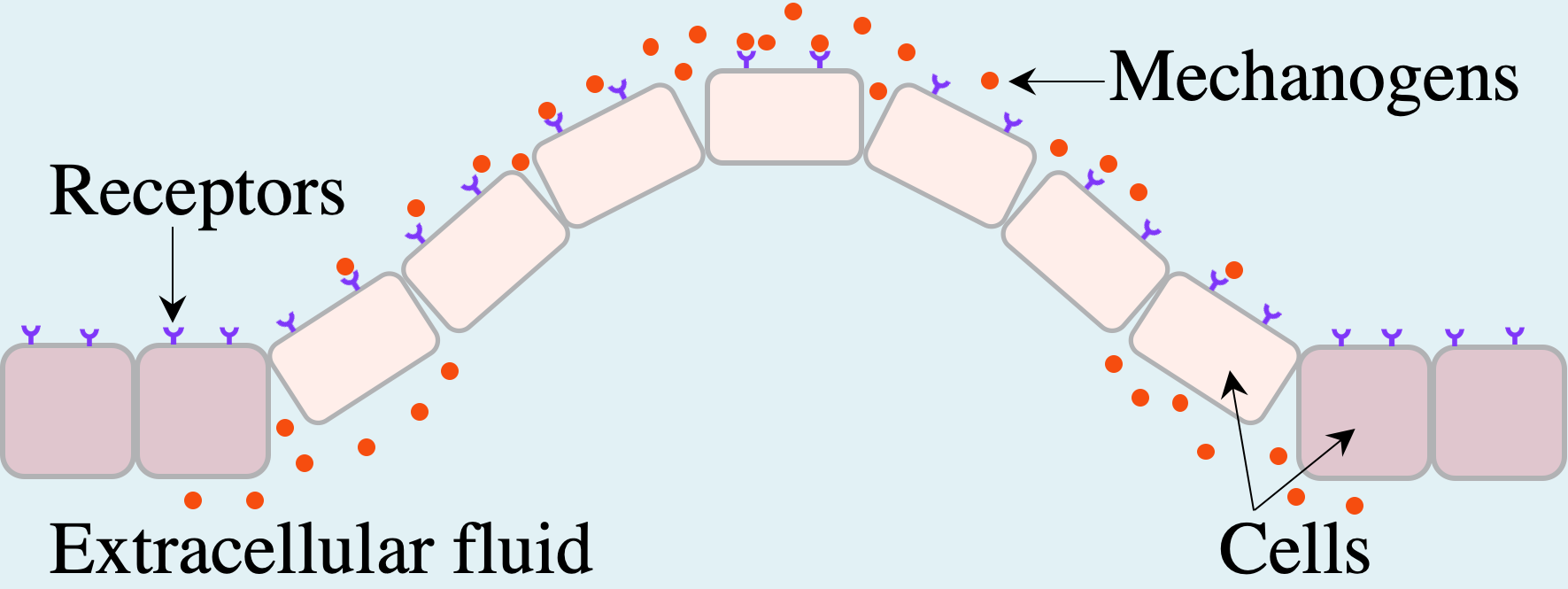}
		\begin{tabular}{cc}
			(b)&(c)\\
			\includegraphics[width=0.24\textwidth]{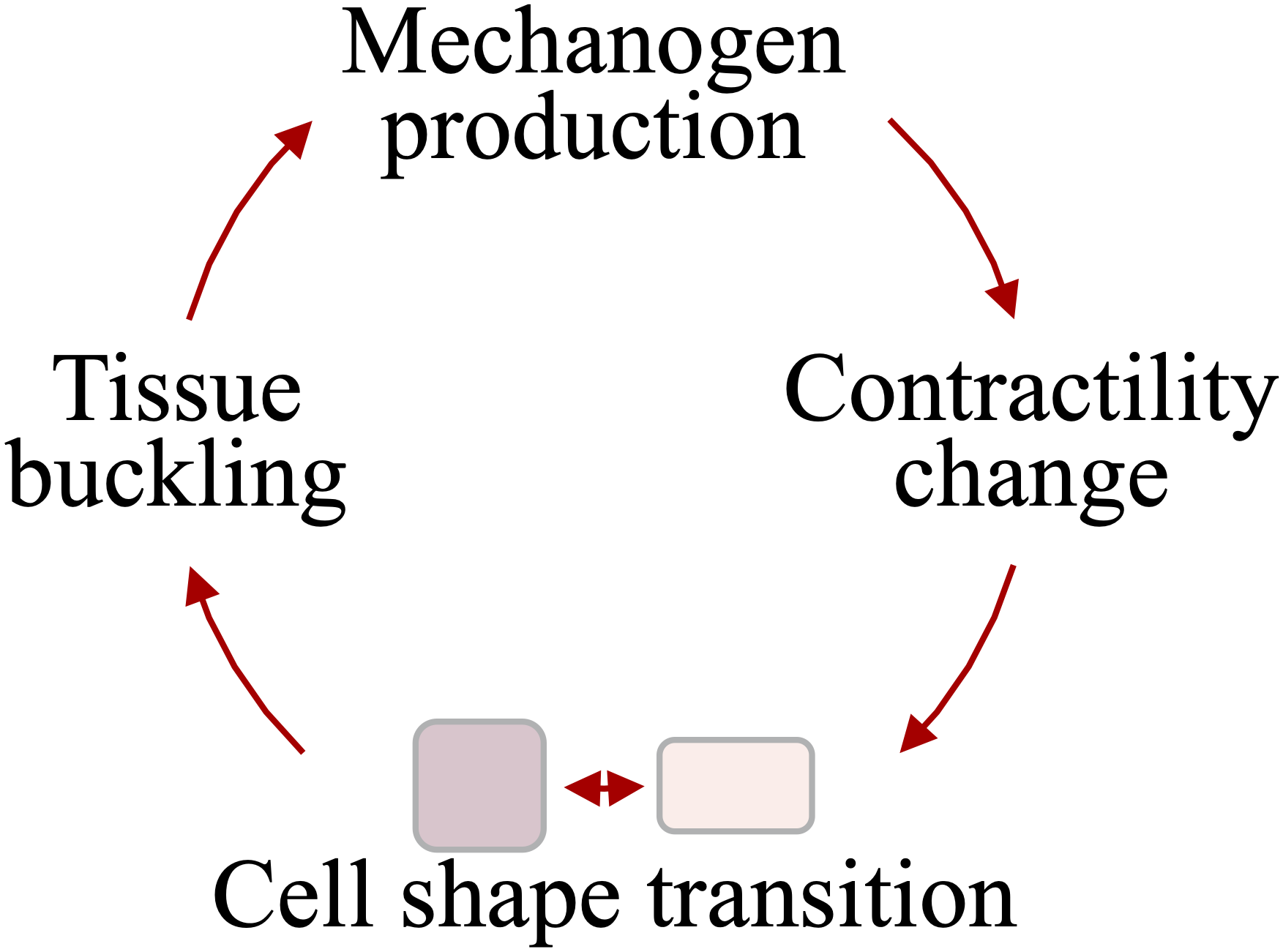}
			&\includegraphics[width=0.22\textwidth]{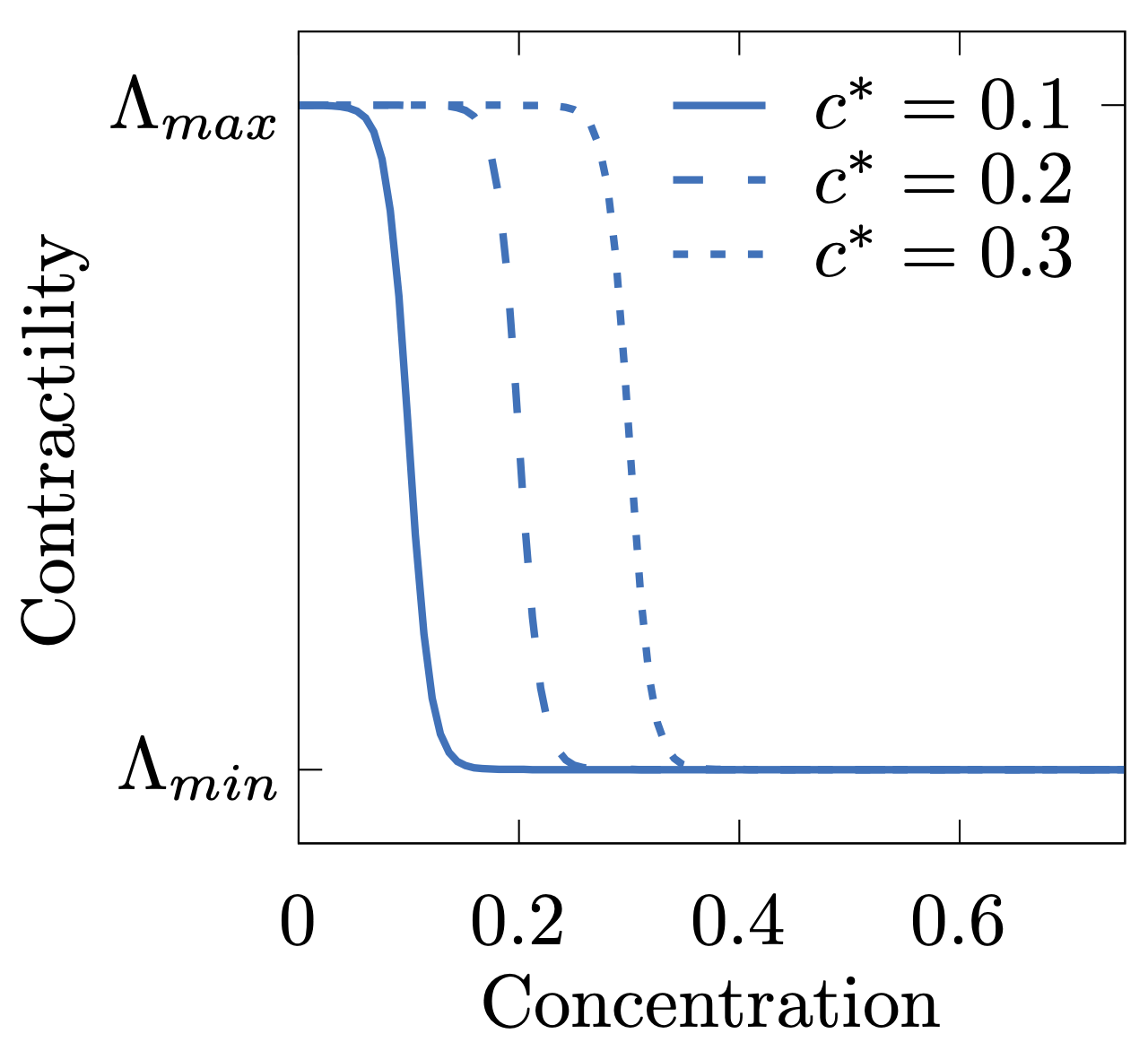}
		\end{tabular}
		\caption{(a) Schematic of  buckling of a tissue sheet in response to chemical signals (mechanogens) that bind to receptors on the apical (upper) surface of cells and trigger change in cell contractility, and therefore cell shape (in this case, the aspect ratios as a result of force balance \cite{hannezo_14}). Outward tissue curvature in turn stimulates the release of mechanogens into the extracellular fluid, where they can diffuse and influence cell contractility.  (b) Schematic of the mechanochemical feedback considered in our model. (c)  Thresholded dependence of cell contractility on mechanogen concentration,$\Lambda(c)$, assumed in Eq.(\ref{Eq_Lambda}), which give rise to domains of differential tension that lead to buckling of the tissue sheet.}
		\label{fig:Schematics}
	\end{figure}

	Inspired by the biochemical patterning of mechanical forces in thin layers of tissue, we introduce a model of chemical-induced pattern formation in thin elastic sheets.  We consider the situation of a thin monolayer of cells tightly adhered to each other and surrounded by extracellular fluid, as shown in the schematic in Fig.\ref{fig:Schematics}a. The cells change their shape in response to diffusible chemical signals (which we term ``mechanogens’’) in the extracellular fluid, that are secreted by the cells themselves, and that bind to receptors on the cells’ apical (top) surfaces  and trigger changes in the contractile tension of their actomyosin cytoskeleton. For concreteness, we consider mechanogens that relax contractility and increase cell apicobasal surface area while reducing their lateral surface area to conserve volume.  
	The resulting spatially inhomogeneous in-plane expansion of the tissue sheet causes the region exposed to higher chemical signal to buckle out of plane. We then explore the role of mechanochemical feedback (Fig.\ref{fig:Schematics}b), that is, the chemical concentration is in turn affected by the tissue mechanical state, specifically its curvature.

	The cell monolayer, that could be freely suspended or adhered to a thin basal substrate, is modeled as a continuum elastic shell with uniform properties throughout its thickness. We assume that over the short timescales of interest, the mechanical response of the sheet to induced stresses is elastic and ignore the viscous dissipation and fluid flows that could result from cell rearrangements and motility. Then the mid-surface of the thin shell can be represented as a two-dimensional surface embedded in 3D space that allows a convenient description for out-of-plane deformations and calculation of the deformed 3D shape. The mechanical free energy of deformed elastic shells is determined by contributions from the stretching energy $\mathcal{U}_s$, which is proportional to the shell thickness, $h$ (here, determined by cell size),  and arises due to in-plane compression or extension of the shell, and the bending energy $\mathcal{U}_b$, which is a result of curvature and scales as  $h^3$ \cite{landau_lifshitz_elasticity}. Since thickness is usually small compared to shell size ($h \ll R$), the stretching energy cost dominates that of bending, meaning that it is more favorable to bending than to stretching. For an isotropic and linear elastic material, the elastic energy of a deformed elastic shell is given by  \cite{landau_lifshitz_elasticity},
	\begin{eqnarray}
		\mathcal{U} = \mathcal{U}_s + \mathcal{U}_b = \frac{Y}{2} \int \varepsilon^{2}_{\alpha \beta} dA +  \frac{B}{2} \int \kappa^{2}_{\alpha \beta} dA, 
		\label{Eq_ElastEnergy}
	\end{eqnarray}
	where $Y=Eh/(1-\nu^2)$ is the 2D stretching rigidity with the elastic Young's modulus $E$ and the Poisson ratio $\nu$, $B=Eh^3/[12(1-\nu^2)]$ is the bending stiffness, $\varepsilon_{\alpha \beta}$ is the in-plane strain, and $\kappa_{\alpha \beta}$ is the curvature strain accounting for out-of-plane displacements.
	The strain $\varepsilon_{\alpha \beta}$ depends on the local contraction or expansion, which in living tissues is caused by actomyosin contractility with an effective contractile tension,~$\Lambda$. For simplicity, we assume a linear dependence of in-plane expansion on contractility change. The curvature strain $\kappa_{\alpha \beta}$ in the bending term arises as a result of the tendency to reduce stretching. A discretized version of Eq.(\ref{Eq_ElastEnergy}), which we use in our numeric simulations, is described in detail in the Appendix \ref{appendix:a}.
	
	Several recent studies have identified that a change in concentration of signaling agents (``mechanogens'') induces change in tissue contractility \cite{gilmour2017morphogen}.
	Thus, it is natural to consider a dependence on concentration $c$ of the tissue surface contractility $\Lambda(c)$, that develops in the actomyosin belt and tends to constrict the cell apical surface. The contractility change in response to the chemical signal could be either positive or negative depending on the specific chemical.  
	
	Additionally, developing tissue is usually subdivided into discrete regions of different cells according to thresholds in concentration of morphogen gradients \cite{goodwin2020leopard}, which lead to the classic "French flag" pattern of gene expression induced by morphogens \cite{wolpert1969positional}. Although such compartmentalization is not general in physical systems, it is natural and common in living organisms. Inspired by patterns in biological tissues, we assume non-linear coupling between chemical pattern and mechanical stress, and impose the response in contractility strength using a sigmoid function,
	\begin{eqnarray}
		\Lambda(c) = (\Lambda_{\textrm{max}} e^{c^\ast b}+ \Lambda_{\textrm{min}} e^{b c})/(e^{c^\ast b}+e^{b c}),
		\label{Eq_Lambda}
	\end{eqnarray}
	where $c^\ast$ defines the concentration threshold at which contractility changes from its maximal to minimal value, $\Lambda_{\textrm{max}}$ and $\Lambda_{\textrm{min}}$, respectively, and $b$ is the steepness that prescribes the width of transitional zone. This gives non-linear dependence of contractility on concentration leading  to sharp interfaces between regions of different cell contractility (Fig.\ref{fig:Schematics}c).   
	The mathematical form of this sigmoid dependence of $\Lambda(c)$ is a specific choice we make to demonstrate our model predictions. A different choice that leads to a similar sharp transition in cell area is expected to lead to qualitatively similar patterns, because mechanical buckling phenomena are generic and do not rely on specific biomolecules. What is ultimately important for the mechanical deformations in our model, which follow from the general theory of thin elastic shells, is the existence of domains of unequal cell surface area, which can arise in practice through different biophysical mechanisms including cell division-induced expansion. Also, differential expansion can arise even if $\Lambda$ changes linearly with concentration, $c$, because cell shape was shown to be a bistable function of $\Lambda$ in Ref.~\cite{hannezo_14}.
	
	Cell shape in living tissue is determined by a balance of mechanical forces arising from actomyosin contractility at both cell apical and lateral  surfaces as well as cell-cell and cell-substrate adhesion \cite{hannezo_14}, all of which can in principle be affected by mechanogens.  Following previous works that assume incompressibility, that is conservation of the cell volume, the local area expansion caused by the chemical is accompanied by a reduction of the thickness. The case of coupling between concentration and lateral, instead of outer surface, tension has similar but reverse effect producing an increase in thickness and surface area contraction with $c$ due to volume conservation. Dependencies of cell shape on chemical concentration alternative to Eq.(\ref{Eq_Lambda}) can therefore occur. However, in scenarios where the cell volume can change or if the tissue retains its thickness and expands in-plane through cell division, the surface contractility, $\Lambda$, remains a convenient effective parameter  that allows  in--plane area changes required for tissue buckling.
	
	On the other hand, it was shown experimentally that production rate of the signaling regulating developmental gene expression is coupled to mechanical stress \cite{desprat2008tissue, brouzes2004interplay, kirby2016stretch}. Similar observations indicate the higher expression of morphogens at regions of high curvature \cite{hobmayer2000wnt}. We close the mechano-chemical feedback loop by assuming that production of chemicals is coupled to the local curvature. Though such a dependency is written on general grounds, we speculate that a plausible mechanism for the curvature-dependent release of chemicals is the opening of gaps between cells when tissue bends as depicted in (Fig.\ref{fig:Schematics}a). Within this assumed model, a curved shape (non-zero mean curvature) allows the extracellular fluid to surround the cells, and, if the cells produce the mechanogen at their lateral surface, to spread it out from the source. Planar shapes keep the  extracellular space closed restricting the transport through the lateral surface. However, if the receptors are located only on one side of cells, apical or basal, then concentration only on the respective side needs to be considered. 
	Recent experiments \cite{shyer2015bending} have demonstrated similar scenario for  morphogen gradients arising in regions of high curvature and spreading only on one side of the epithelial layer due to the difference in morphogen-receptor interactions on different sides. We also assume the usual uptake of mechanogens which leads to their removal from the extracellular space leading to a linear degradation rate.  The mechanogens can be transmitted along the surface of tissue or diffuse in the extracellular fluid, forming  concentration gradients. The time evolution for a chemical signal is then governed by the equation
	\begin{eqnarray}
		\partial_{t} c &= D\nabla^2 c +\frac{\Theta(H)w H}{1+ H/H_{\textrm{s}}}-\beta c,
		\label{EqC}
	\end{eqnarray}
	where $D$ is the diffusivity, $H$ is the mean curvature with the Heaviside function $\Theta(H)=1$, when the curvature radius is oriented outward and $\Theta(H)=0$ when inward, which arises from  differences between apical and basal surfaces; 
	$H_{\textrm{s}}$ is a characteristic curvature at which the chemical production saturates to a rate given by $w H_{\textrm{s}}$, and $\beta$ is the rate of degradation. The maximum local steady state concentration at a region of high curvature, $H \gg H_{\textrm{s}}$ is then given by $c_{\textrm{max}} = w H_{\textrm{s}}/\beta$. 
	
	The chemical kinetics given by the chemical diffusion,  production and degradation rates defined in Eq.(\ref{EqC}) are typically slower than elastic stress propagation.  For constant production rate, the timescale to reach a steady state in concentration is given by $\beta^{-1}$, which for morphogens in developing tissue has been measured to be of the order of \emph{hours} \cite{kicheva_12}. This is at least one order of magnitude slower than the time scale for contractility and tissue length remodeling in epithelial cell sheets, which is of the order of \emph{minutes} \cite{cavanaugh2020rhoa,martin2009pulsed,choi2016remodeling}.
	The strain relaxation by out-of-plane deformations is assumed to be the fastest process in the mechano-chemical loop we consider. We assume that the elastic energy relaxes through the overdamped dynamics given by, $\gamma \partial \mathbf{u}/\partial t= - \delta \mathcal{U} / \delta \mathbf{u}$, where $\mathbf{u}$ is the material displacement. We consider viscous damping by the surrounding fluid and neglect any possible viscous remodeling of the tissue material \cite{Matoz-Fernandez2020}. For a sheet of characteristic size, $R$, immersed in a fluid of viscosity, $\eta$, the frictional drag on the sheet goes as $\gamma \sim \eta R$. A small out-of-plane deformation $\delta x$ leads to a bending elastic restoring force that can be estimated from Eq.(\ref{Eq_ElastEnergy}) to be $ E \cdot h R^2 \cdot h^2 (\delta x/R^2)^2 \cdot \delta x^{-1} \sim E h^3 \delta x/R^2$. By balancing this against a viscous drag force of $\eta R \delta x/ \tau_{\textrm{el}}$, we can estimate a characteristic timescale for the relaxation of the elastic deformation energy, $\tau_{\textrm{el}} \sim (\eta/E) \cdot (R/h)^3$. For the typical material properties of a suspended epithelial monolayer  \cite{Harris2012} ($E \sim 10$ kPa, $\eta \sim 1$ Pa$\cdot$s, $R/h \sim 10^2$) we get $\tau_{\textrm{el}} \sim 10^1 -10^2$s.
	
	Finally, the feedback loop has a coupling between tissue deformations and the chemical production rate. Since the production rate is associated with opening interstitial gaps, we assume a very small delay in this process, but the following chemical rearrangements are slowed down due to limited diffusion discussed above. Thus, slow modes of chemical rearrangements evolve on a longer time scale, while fast modes of mechanical conformations follow them quasi--statically, accordingly to Haken's slaving principle: ''fast modes are slaved by slow modes'' \cite{haken1983introduction}. This separation of timescales allows us to reproduce the tissue morphing dynamics in simulation by implementing an iterative procedure. Starting with the reference equilibrium configuration, a small perturbation in concentration is introduced, which leads to a local contractility change. 
	The stress is updated according to the altered contractility, and since the mechanical response is fast, we find actual three-dimensional shapes by minimizing the elastic energy Eq.(\ref{Eq_ElastEnergy}). As the system reaches a mechanical equilibrium, the concentration profile is updated by a small time step taking into account the new shape that prescribes a new production rate. The iterative process continues till the change in concentration profile, which is coupled with reshaping, vanishes.  Since the pattern develops on a domain of finite size, the dynamics slows down and eventually becomes stationary.  
	
	We also tested the emergent dynamics and patterns assuming comparable timescales for mechanical remodeling and chemical rearrangements. In this case we performed a steady-state dynamics by  sequential computations of mechanical and chemical equilibrium states. Iteratively repeating the update, the system dynamics eventually converged. This approach is more computationally time-consuming and does not allow to reproduce dynamics with a time scale related to a measurable parameter. However, it still captures the same resulting configurations as for computations when we assume separation of time scales described earlier. This demonstrates that even if our assumption of separation of time scales is not true in a given experimental system, the approach we use captures the final steady states.
	
	\section{Results}\label{Section_Res}
	
\begin{figure*}[t]
		\centering
		(a)\\
		\includegraphics[width=0.875\textwidth]{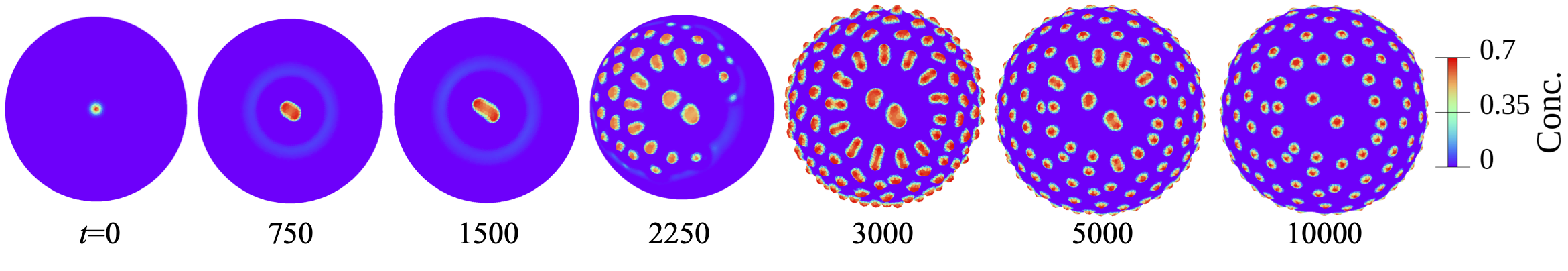}\\
		(b)\\
		\includegraphics[width=0.95\textwidth]{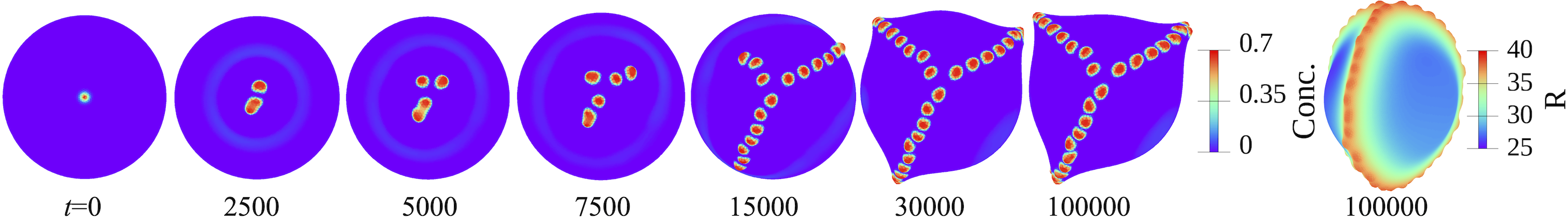}
		\caption{Evolution of propagating chemical patterns and equilibrium shapes of a spherical shell at thickness $h=2.0$, diffusion $D=0$, and different thresholds (a) $c^\ast=0.1$, and (b) $c^\ast=0.2$. Concentration fields are shown in top view while the most right panel for the radius color map in (b) is in a side view. Simulation snapshots demonstrate occurring elastic instabilities: buckling to a bulge at the initial perturbation resulting in increased concentration and widening the spot ((a) $t=750$), bulge splitting ((a) $t=1500$ and (b) $t=2500$), buckling at secondary ring ((a) $t=750$ and (b) $t=2500$) and wrinkling along the ring ((a) $t=2250$). The ridge-like propagating pattern (b) arises at high $c^\ast$ when curvature at the secondary ring does not cause enough production to exceed the threshold of contractility change, whereas it becomes spotted (a) at low $c^\ast$ by means of wrinkling at the secondary ring.} 
		\label{fig:SphereD0evo}
	\end{figure*}

	Consider a spherical shell of uniform radius $R$ and small thickness $h \ll R$ in the undeformed state, representing a developing tissue sheet.  Assume the tissue remains at a spontaneous curvature, $1/R$, without any residual, in--plane stress. Thus, without any further perturbation, the total elastic energy of the spherical shell is vanishing and it remains in a stationary, mechanical equilibrium state. Our assumed natural curvature does not necessary imply wedge-shaped cells, but the intercellular gaps have to be initially non-uniform over the thickness in this case.
	Then we introduce an initial small point perturbation of the chemical profile defined by a Gaussian function with the spot size comparable to the shell thickness, $d=h$, and magnitude of the  order of the upper concentration limit $c = c_{\textrm{max}}$ ( Fig.\ref{fig:SphereD0evo} at $t=0$), causing a local point--like inhomogeneity in contractility. 
	The shell now has two domains: an outer one with a higher initial contractility ($\Lambda_{\textrm{max}}$), and a smaller, inner one with higher chemical concentration and reduced contractility ($\Lambda_{\textrm{min}}$) that is associated with in--plane expansion and tendency to develop larger surface area.  Thus, being constrained by the outer domain, the inner region is under compressive stress.  With the shell  experiencing  in--plane incompatibility, it deforms to reduce stretching and is prone to bending. The macroscopic manifestation of this incompatibility is out-of-plane deformations of the shell due to buckling instability. The dynamics of deformations depends on mechanical properties of the sheet, coupling between concentration and contractility, and the diffusivity of signaling species. A full list of the parameters we use in our simulations is recapitulated in Table \ref{table:1}. We find that the resulting pattern is sensitive to varying thickness $h$, diffusion $D$, and the concentration threshold, $c^\ast$, at which contractility transitions to its lower value. The threshold $c^\ast$ determines the nonlinear mechanical response that links chemical inputs to shape change actuation. Varying this parameter leads to qualitatively different shapes.
	
\begin{figure}[t]
		\centering
		\includegraphics[width=0.45\textwidth]{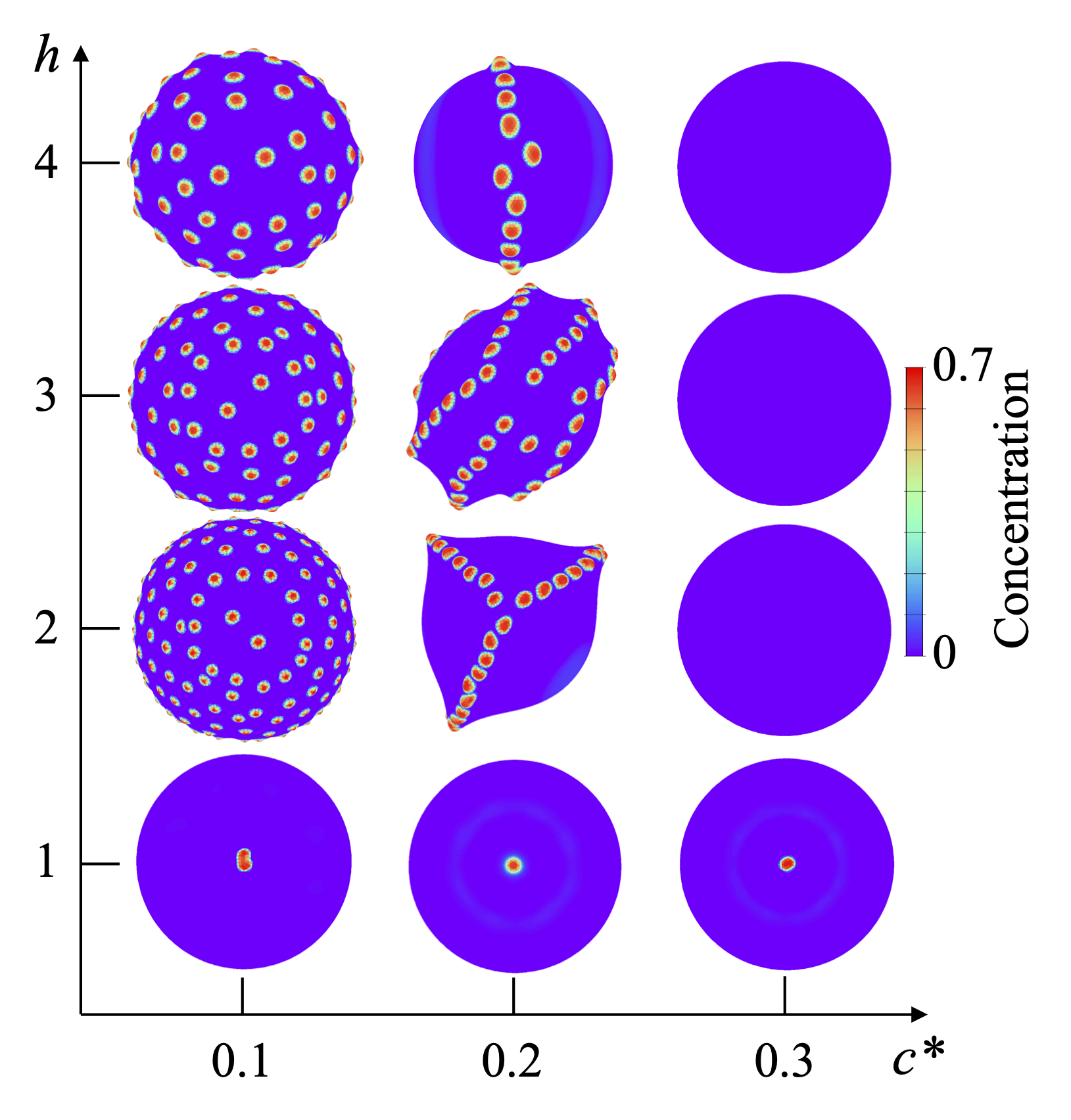}
		\caption{Steady states of a spherical shell colored according to the local concentration at different threshold $c^\ast$ and initial thickness $h$. Patterns emerge as a consequence of a small initial chemical perturbation at a pole of the sphere (the central point in the top views). Higher concentration regions are associated with greater positive (outward) curvature.}
		\label{fig:SphereD0}
	\end{figure}

	We first aim to explore the feedback between curvature and local chemical production as a cue for spatial patterning. To see this, we assume vanishing diffusion $D=0$, such that production is balanced by linear degradation locally. The simulated spheres at $R=30$ reveal that, even in the absence of diffusion, coupled mechanical deformations and chemical production result in propagating patterns, as shown in Fig.\ref{fig:SphereD0}. We find that when thickness $h>1$ and threshold $c^\ast<0.3$, the initial excitation makes the state unstable and the shape becomes distorted via buckling instability. At small $h$, the shell can develop larger amplitude out-of-plane deflections for the same bending energy cost \cite{timoshenko2009theory}. In this case (lower row in Fig.\ref{fig:SphereD0}), only a single localized bulge develops in the inner domain of lower contractility, while the outer region with the larger contractility retains its shape. Increasing thickness, $h$, leads to smaller out-of-plane deflections at a given compressive load  leading to a smaller curvature and longer length scale deformations that extend out of the initial perturbation spot \cite{timoshenko2009theory}. In this case, the positive mechanochemical feedback allows the pattern to propagate because the deflected region has higher curvature which leads to chemical production, which in turn drives the incompatibility responsible for out-of-plane deflection.

	For lower concentration threshold, $c^\ast=0.1$, there is more mechanical stress and incompatibility since the inner region over which the shell relaxes its contractility is larger. In this case, the shell adopts a pattern of almost uniformly distributed spots of high concentration, where each spot locally deforms the shell to a bulged shape. I moved the next sentence here. The spot size and separation between spots is seen to increase with $h$. The scaling law and underlying mechanism is discussed in Sec. \ref{Section_Buckling}. At intermediate $c^\ast=0.2$, the emergent bulges form ridge-like lines that propagate in random directions if the initial perturbation is symmetric, but demonstrate a preferred direction when the initial spot is elongated due to non-axisymmetric perturbation. 
	In case of a high $c^\ast>0.3$, the shell is stable to the initial perturbation because the buckling-induced curvature developed at a given thickness does not cause enough chemical production required for contractility change. 
	In this case, the chemical concentration level eventually decreases at steady state due to linear degradation in Eq.(\ref{EqC}). The production term becomes dominant when mean curvature $H>\beta c/(w-\beta c /H_s)$, which corresponds $H>0.03$ at given $\beta,w,H_s$ (Table \ref{table:1}) ultimately leading to a chemical concentration higher than the threshold  $c^\ast=0.1$.
	However, if $H$ does not develop this critical value,  and  if $c^\ast$ is high, not enough chemical is produced away from the spot to propagate the pattern. Thus, the small initial  fluctuation decays, making contractility uniform over the shell. 
	
	The time series for patterns of two representative cases are depicted in Fig.\ref{fig:SphereD0evo}. A typical evolution in both cases starts with a small perturbation in the concentration that leads to a local decrease in contractility followed by area expansion and buckling. As curvature increases with buckling, it causes production enhancement, and as a result the size of the initial bulge grows. Then a ring of higher concentration is formed at a short distance from the initial spot ($t=750$ and $t=2500$ in Fig.\ref{fig:SphereD0evo}a,b, respectively), which at low threshold $c^\ast=0.1$ breaks into multiple bulges along the ring ($t=2250$ in Fig.\ref{fig:SphereD0evo}a). Each high concentration spot becomes wider following the same dynamics as the initial bulge, and creates another row of bulges. The process eventually slows down and  terminates when the pattern occupies the entire spherical shell, with uniformly distributed spots ($t=10000$ in Fig.\ref{fig:SphereD0evo}a), except at the site of the initial imperfection. 

	In the case, $c^\ast=0.2$, the ring formed around the initial perturbation does not generate a contractility change in that region. 
	For a given thickness, the concentration at the secondary ring is constant at different $c^\ast$, but spots emerge only at a low threshold.  Thus, when the ring does not break into multiple spots, an initial bulge continues instead to split into two separate bulges ($t=2500$ in Fig.\ref{fig:SphereD0evo}b) and this division propagates along the shell creating ridge-like deformations. The lines of bulges retain their position and strongly distort the spherical shell shape when compared with the  uniformly distributed spots at a lower $c^\ast$.
	
	Although the bulges can be organized into very different patterns, the common feature of both types of patterns is the formation of separate bulges. The reason for bulge formation is the high mean curvature when the shell develops cap-like shapes with positive curvature in two orthogonal directions coupled to higher local production of chemical at these bulges. Wrinkles or folds with curvature only in one direction, provide a weaker feedback and do not cause contractility change, thus bulges become eventually round-shaped even if they were elongated at earlier stages of evolution ($t=3000$ in Fig.\ref{fig:SphereD0evo}a). 
	
\begin{figure}[t]
		\centering
		\includegraphics[width=0.475\textwidth]{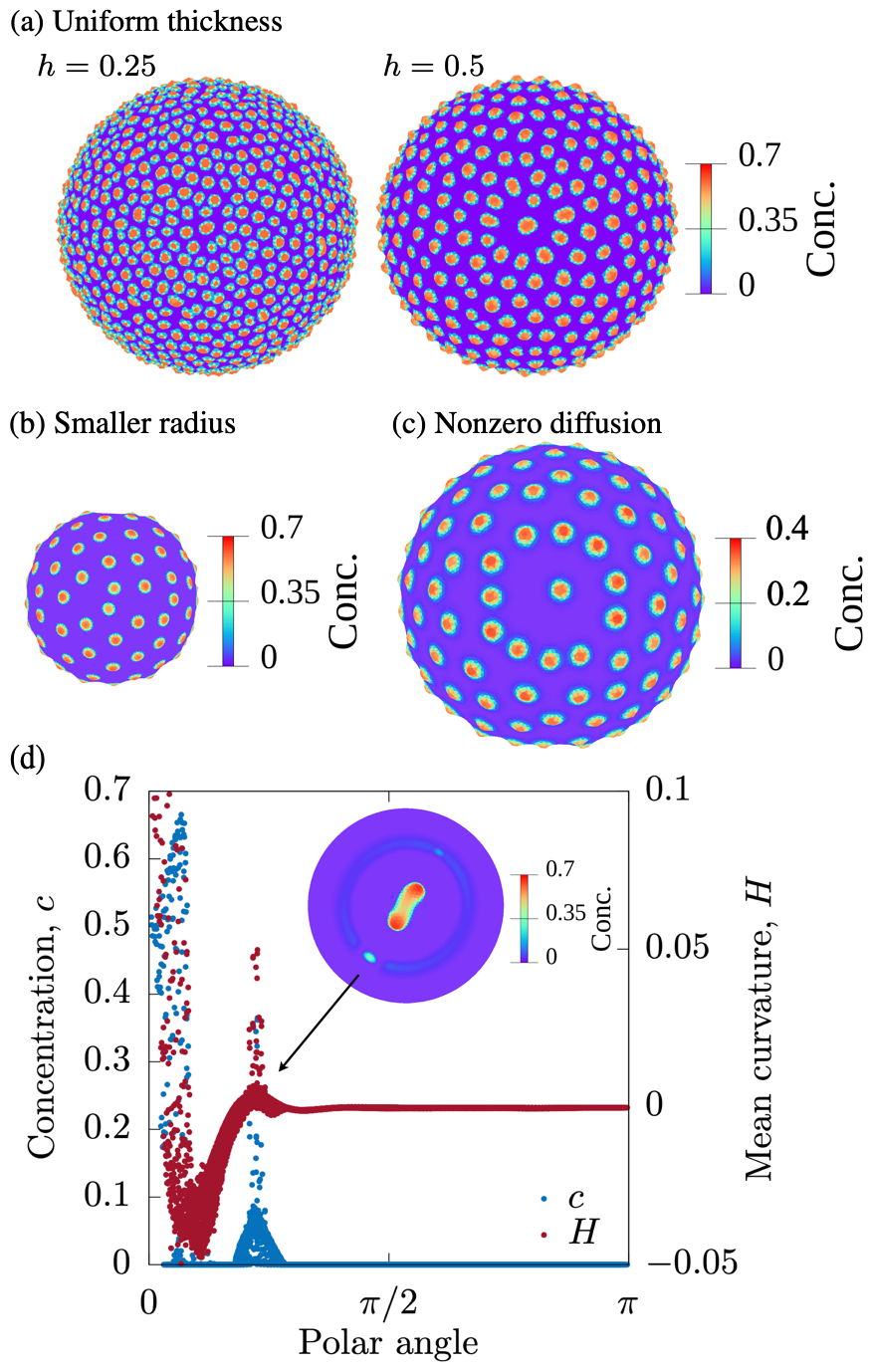}
		\caption{(a) Chemical patterns arising on a spherical shell of fixed uniform thickness $h$ at $D=0,  c^\ast=0.1, R=30$. Shells obeying the volume conservation condition at (b) smaller radius $D=0, h=2, c^\ast =0.1, R=15$ demonstrate the same patterns obtained at large $R$, while (c) increasing diffusion leads to a larger spot size $D=1.5, h=2, c^\ast =0.1, R=30$, than seen in Fig.\ref{fig:SphereD0evo}a. (d) Concentration and mean curvature at nodes of mesh on sphere depending on polar angle with the initial perturbation at the pole. The secondary peak at a nonzero polar angle represents the secondary ring that forms as a result of the buckling instability around the initial pertubation at the pole.}
		\label{fig:unih}
	\end{figure}

	The assumption of cell volume conservation causes the shell to become thinner in regions where in-plane contractility is lower. We now show that even if cell volume is not conserved and the shell remains at uniform thickness, similar spotted patterns can occur.  In Fig.\ref{fig:unih}a, we display stationary shapes for spheres with $c^\ast=0.1$ and of uniform thickness (shown for two different shells with varying thickness, $h$). We notice that the pattern can now be propagated at smaller thickness for the same contractility change. This is because the inner region with $\Lambda_{\textrm{min}}$ remains at the initial thickness and is therefore thicker now than the corresponding volume-conserved case. This leads to smaller out-of-plane deflection and curvature and the critical initial shell thickness at which propagating patterns result is lower than for the volume-conserved shells presented before.

	The simulated spheres of different radius reveal that spot separation and spot size are independent of the sphere radius. In Fig.\ref{fig:unih}b, we demonstrate an example of a sphere at $R=15$ accommodating proportionally smaller number of spots compared to the previously discussed cases with $R=30$, showing that the spot spacing is independent of shell size. This is because we assumed the shell has an intrinsic curvature $1/R$ and no residual bending stress.  However, the patterning is expected to be strongly affected by the sphere radius if this condition is not satisfied. In such cases, the production rate in the initial state will be non-zero and increase with $1/R$, and thus, Eq.(\ref{EqC}) will be unbalanced.   
	
	So far in this paper, we have considered  pattern formation at vanishing diffusion rate valid for slow chemical spreading. For a single-cell layer of  epithelium tissue, the typical thickness is $ h \sim10\si{\um}$, which corresponds to a unit of length in our simulations. Thus, taking into account that the upper limit for experimentally measured diffusion coefficient of morphogens in extracellular fluid and in tissue is $\sim 50 \si{\um^2/s}$ \cite{kicheva_12}, the maximum reasonable value in simulations is $D \approx 0.7$ . By performing simulations at a different $D$ and keeping other parameters constant, we found that the patterns, which appear at $D>0$, show only quantitative change compared to $D=0$. In Fig.\ref{fig:unih}c we demonstrate the concentration field and 3D shape at large diffusion $D=1.5$, which significantly exceeds the diffusion rate in real biological systems. One can see that spot size is enlarged even though the maximum concentration decreases due to spreading. However, an increased diffusion $D>0.4$ leads to a propagating pattern with multiple spots at $h=1, c^\ast = 0.1$, while it remains in the single bulge equilibrium state at $D=0$ (bottom left case in Fig.\ref{fig:SphereD0}), thus, providing a qualitatively different shape. As expected, diffusion helps to scale up spot size and propagate the pattern by spreading the chemical more uniformly over the surface.
	
	Note that without coupling between deformations and chemical production, the initial perturbation does not lead to a propagating pattern. The feedback is clearly seen in Fig.\ref{fig:unih}d, where higher mean curvature, $H$, (red data points) is followed by increased concentration (blue data points). One can see that concentration significantly exceeds the threshold $c^\ast=0.1$ in the vicinity of the initial perturbation at small polar angle and then decreases as curvature $H$ becomes negative. A small increase in $H$ at a larger angle leads to the secondary peak in $c$ and the formation of bulges at small $c^\ast$ (the inset to Fig.\ref{fig:unih}d). However, this secondary peak does not cause a contractility change at higher threshold, \emph{e.g.} $c^{\ast} = 0.2$, for which $c \ll c^\ast$, as seen in Fig.\ref{fig:SphereD0evo}b. If the production term in Eq.(\ref{EqC}) is independent of the shape change, the concentration will decrease with time due to a linear degradation, followed by decreasing domain of low contractility, and eventually the initial buckling will disappear. Also, in the case of a constant source of chemicals without feedback, the system will approach a steady state resting at a balance between production and degradation.  This last scenario will also give rise to a pattern with wrinkles in the expanded domain. The wrinkles will decay and will have increasing wavelength away from the interface, leading to a qualitatively different type of pattern \cite{vandeparre2011wrinkling}. 
	
	\section{Analysis of Elastic Instabilities}\label{Section_Buckling}
	
	We now seek insight into the mechanism of formation of the patterns shown in Fig.\ref{fig:SphereD0} and its dependence on concentration threshold and shell thickness. In particular, we would like to understand why the spotted pattern arises at small $c^\ast$ whereas the initial bulge subdivides into two separate bulges forming the linear ridge pattern at intermediate $c^\ast$,
	
	We distinguish three different elastic instabilities that occur successively in the shell. First, the initial perturbation in chemical field associated with the contractility change leads to buckling and formation of the initial bulge that attains different shapes depending on $c^\ast$ ($t=750$ in Fig.\ref{fig:SphereD0evo}a and $t=2500$ in Fig.\ref{fig:SphereD0evo}b). At the same time, a ring of larger curvature and concentration appears at a short distance from the initial bulge. Finally, for the spotted patterns occurring at low $c^{\ast}$, this secondary ring breaks into multiple equally separated bulges ($t=2250$ in Fig.\ref{fig:SphereD0evo}a). 

	\begin{figure}[t]
		\centering
		\includegraphics[width=0.475\textwidth]{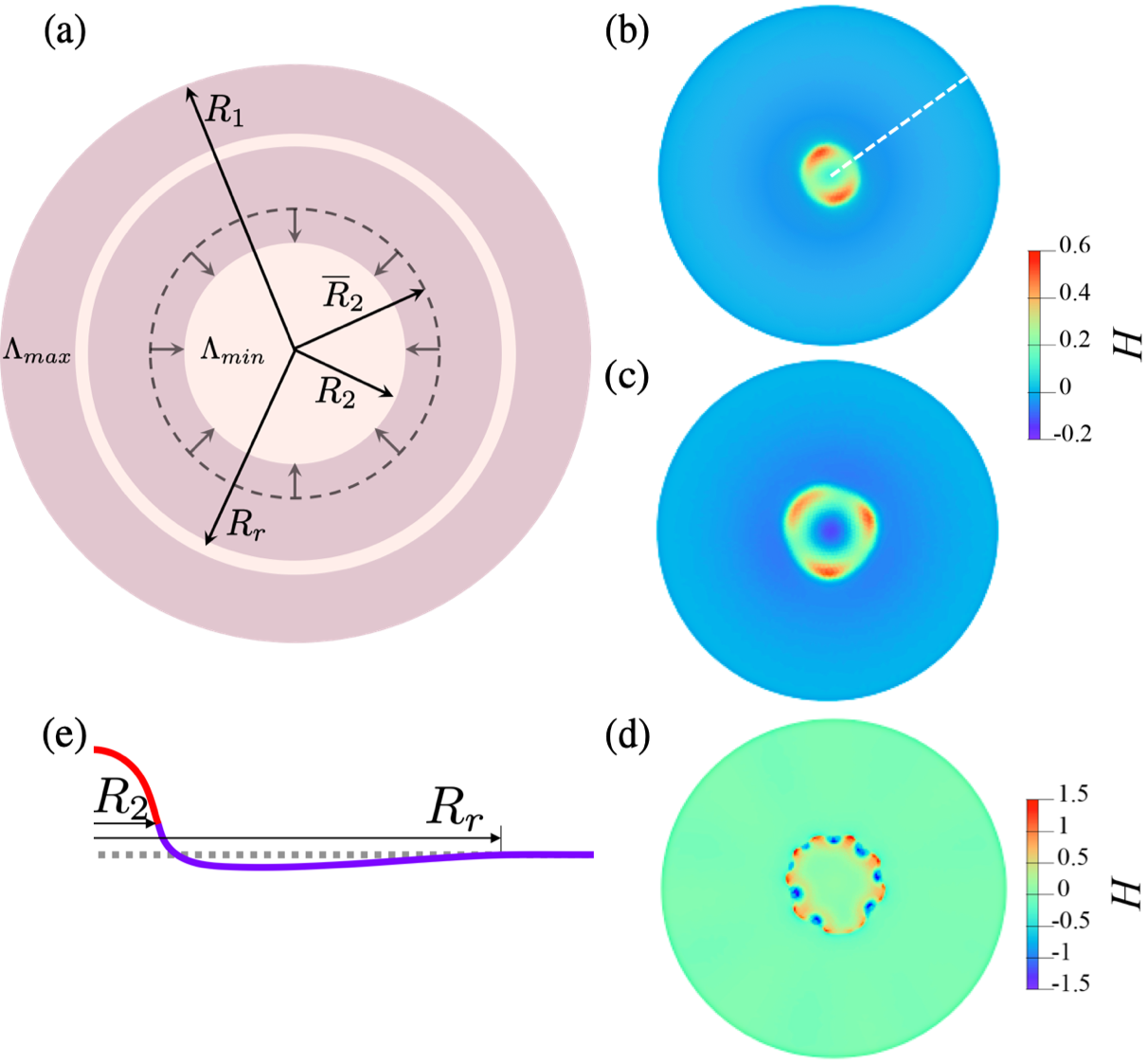}
		\caption{ (a) Scheme of buckling instabilities for the initial bulge and the secondary ring. (b) Wrinkling instability in a disk of contractility $\Lambda=15$  with the inner circle of lower contractility $\Lambda=10$ and decreased thickness demonstrating different number of wrinkles depending on the initial thickness and size of the inner circle at (b) $h=2$, $R_2=2$, (c) $h=2$, $R_2=3$, and (d) $h=0.25$, $R_2=3$.  
			(e) The profile of the shell middle line in the vicinity of inner circle (only one half is shown due to symmetry) along the radial direction  depicted by the white dashed line in (b) where the difference between deflections in two domains is comparatively small. Dotted line is the initial planar shape, while it develops a spherical cup shape in region with low contractility (red line) and small out-of-plane displacements in the outer region (purple line), which bring about the appearance of the secondary ring of radius $R_r$.
		}
		\label{fig:Buckling}
	\end{figure}

	Let us examine the instability that occurs first, which causes the initial bulge on the shell. This is also similar to the third instability in the sequence described above, where the secondary ring develops wrinkles. Since we assumed that the spherical shell has no bending in initial state, and showed that the curvature of the sphere does not affect the pattern, we can obtain the same pattern on a piece of the spherical shell assumed to be nearly flat. For simplicity and in order to focus on the mechanical deformations alone, we numerically test the elastic instabilities on a disk of radius $R_1$ and initial thickness $h$. While the  disk in an undeformed, reference state has initial uniform contractility $\Lambda_{\textrm{max}}$, we consider the deformations induced by an inner, circular domain of radius $R_2<R_1$ whose contractility is lowered to $\Lambda_{\textrm{min}}$ , as depicted in Fig.\ref{fig:Buckling}a, mimicking the effect of a higher chemical concentration in the initial spot. Similar to simulations on a sphere with natural curvature $1/R$, we assume that the sheet is flat in the initial state, that its elastic properties are isotropic, and that the contractility is constant in time. By prescribing values of $\Lambda_{\textrm{max}}$ and $\Lambda_{\textrm{min}}$ we avoid solving the chemical concentration updates in Eq.(\ref{EqC}), and perform minimization of the elastic energy Eq.(\ref{Eq_ElastEnergy}) to investigate the buckling processes involved in the formation of the initial pattern.  The inner circle expands laterally in both radial and circumferential direction due to the lower contractility $\Lambda_{\textrm{min}}$, as it tries to reach an optimal radius $\overline{R}_2={R}_2(\Lambda_{\textrm{max}}/\Lambda_{\textrm{min}})$, and decreases its thickness to a value $\overline{h}$ due to the volume conservation. It is also confined by the outer region that does not change its area and remains at the initial thickness. The biaxial compression caused by this constraint can be fully or partly removed by buckling. The buckled shape depends on the elastic and geometric properties of the shell and in an equilibrium state satisfies the minimum of elastic energy Eq.(\ref{Eq_ElastEnergy}).
	
	Deformations caused by confinement in the radial direction alone can be described as the classical Euler buckling of a plate in response to an in–plane stress in one direction. In this case a plate buckles to a spherical shell, when compressive stress exceeds the critical value \cite{timoshenko2009theory}  $\sigma_{\textrm{cr}}=K^2 B/(\overline{R}_2^2 \overline{h})$, where $K=2.3$ is a geometric constant for the first buckling mode when ends of the plate are not allowed to rotate. Using  $R_2=1, h=2$ and the relation between biaxial stress and strain, we obtain a critical strain $\varepsilon_{\textrm{cr}} \approx 0.1$, which is much smaller than the strain $\varepsilon=\Lambda_{\textrm{max}}/\Lambda_{\textrm{min}}-1=0.5$ exerted due to the contractility change in an unconstrained shell, and thus, the inner circle  exhibits buckling.
	However, it was demonstrated that the buckling of a circular plate under biaxial compressive load leads to more complicated morphologies due to compression in circumferential direction \cite{ortiz1994morphology,holmes2008crumpled}. Several wrinkles arise along the interface between two domains, demonstrating non-axisymmetric buckling modes as $\sigma_{\textrm{cr}}$ increases, while it remains spherical at the lowest eigenmode with smooth deflections and nearly circular boundaries at low $\sigma_{\textrm{cr}}$ value. 
	
	The characteristic wrinkling wavelength, $\lambda$, of an axially compressed, suspended, thin sheet is given by the scaling relation $\lambda \sim h^{1/2}/\varepsilon^{1/4}$, that arises from the minimization of elastic energy \cite{Cerda2003}.  To examine how the wrinkling within the expanded inner region in our setup scales with its size and the thickness of the disk, we study the equilibrium shape of disks with varying $R_{2}$ and $h$.
	Simulated disks of initial thickness $h=2$ with an inner circle of reduced contractility which leads to biaxial expansion by $\varepsilon=0.5$, and a smaller thickness $h/(\varepsilon+1)^2$ to satisfy the volume conservation condition, develop the first wrinkling instability in the azimuthal direction at $R_{2}=1.95$ with $\lambda \sim 6.12$. By varying $R_{2}$, we find that the subsequent transitions to three and four wrinkles occur at $R_{2}=2.9$ and $R_{2}=3.9$, respectively, corresponding to the same characteristic wrinkling wavelength, $\lambda$. This is consistent with the theory expression from \cite{Cerda2003} and with our simulations on a spherical shell, reproducing the same bulge shape ($t=750$ in Fig.\ref{fig:SphereD0evo}a). Two typical examples of buckling instability resulting in different number of wrinkles with increasing $R_{2}$ are depicted in Fig.\ref{fig:Buckling}b-c. Decreasing $h$ at constant $R_{2}=3.0$ leads to increased number of wrinkles with wavelength scaling as $\lambda \sim h^{1/2}$ in agreement with Ref.\cite{Cerda2003}.  To illustrate this, we show the disk shape at a smaller $h=0.25$ in Fig.\ref{fig:Buckling}d, where eight wrinkles of a larger curvature appear near the boundary of the inner circle that relax toward the center.  At the parameters we use to obtain the results reported on a spherical shell in Fig.\ref{fig:SphereD0evo}, two wrinkles tend to form within the initial central bulge.
	Since the initial perturbation in simulations on a sphere is chosen to be small with width of order $h$, this first leads to buckling with no azimuthal wrinkles, and than two wrinkles appear when the bulge increases in radius. Each wrinkle generates increasing concentration due to higher curvature along it, which in turn causes bulge division by developing positive Gaussian curvature in two spots at the wrinkle tips instead of the single initial one.
	
	Buckling and transition to the bulge shape causes propagating deformations in the outer region generating curvature of both signs (Fig.\ref{fig:Buckling}e). Near the interface, the shell has inward deflections to accommodate the gentle slope in the transition region between the thin bulge region and thick planar domain, and it also has a small positive jump in curvature exhibiting oscillatory behavior along the radial direction on a disk or along the polar direction on a sphere  (Fig.\ref{fig:unih}d). Unlike the simple Euler buckling of a slender column, the out-of-plane deflections must be small because they are associated with stretching or compression of the shell in the azimuthal direction, and are thus constrained and oscillate about the initial state leading to the secondary ring. The position of the secondary ring, $R_r$ does not define the final spot separation but slightly disturbs the spot distribution around the initial perturbation, as seen in Fig.\ref{fig:SphereD0} at $c^\ast=0.1$. 
	
	The region of positive curvature forming a ring in the outer domain leads to a higher production of chemical and subsequent contractility change if $c>c^\ast$, which can be satisfied only at small $c^\ast$. The resulting local expansion of the sheet  causes, in turn, another shape instability due to a local biaxial expansion at the secondary ring. Thus, multiple wrinkles appear along the ring to satisfy the incompatibility and forming the next ring at a larger polar angle.  These wrinkles arising within the expanded ring break into bulges that are seen as spots in the final pattern.

	\begin{figure}[t]
		\centering
		\begin{tabular}{cc}
			(a)&(b)\\
			\includegraphics[width=0.235\textwidth]{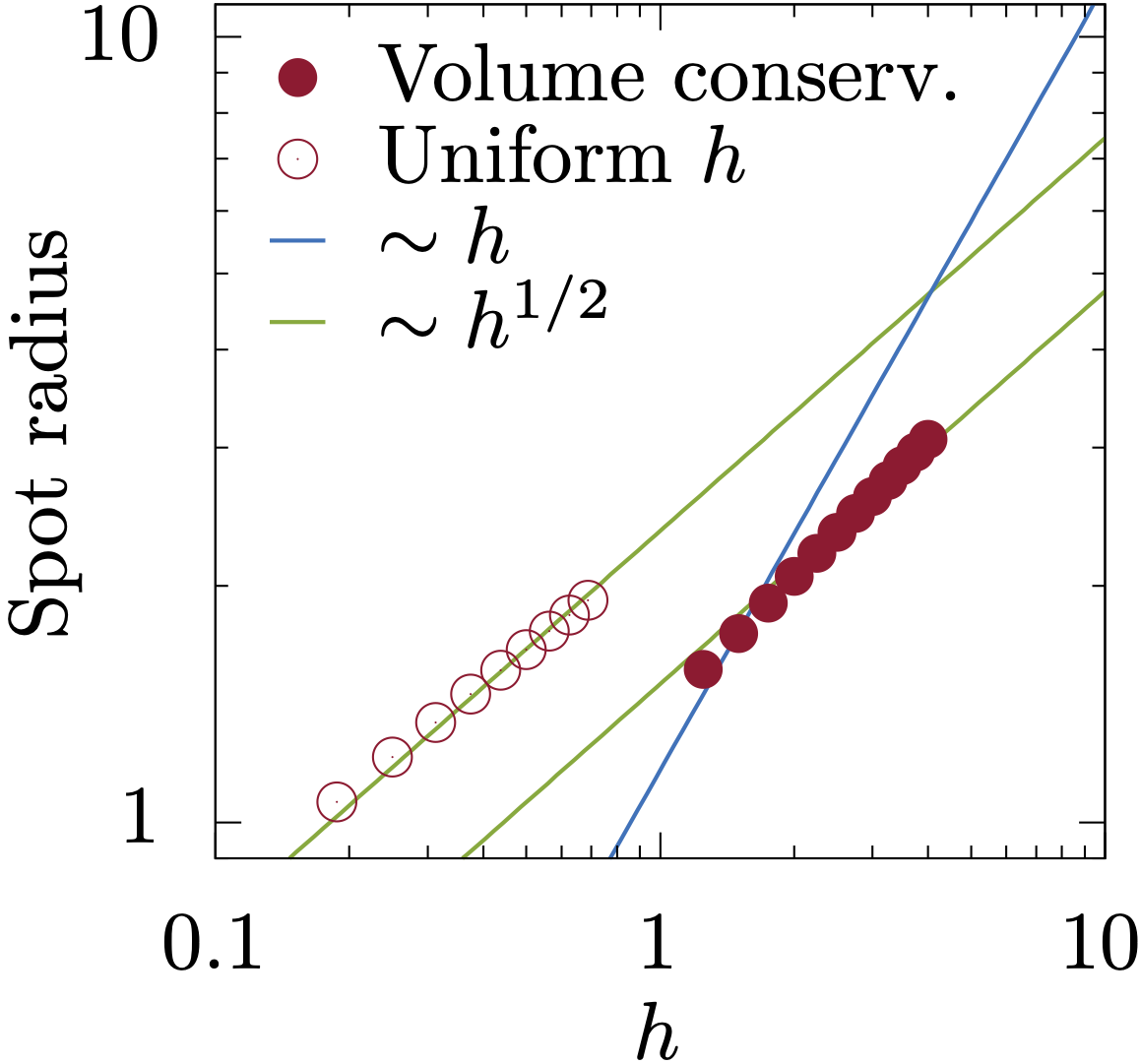}
			&\includegraphics[width=0.235\textwidth]{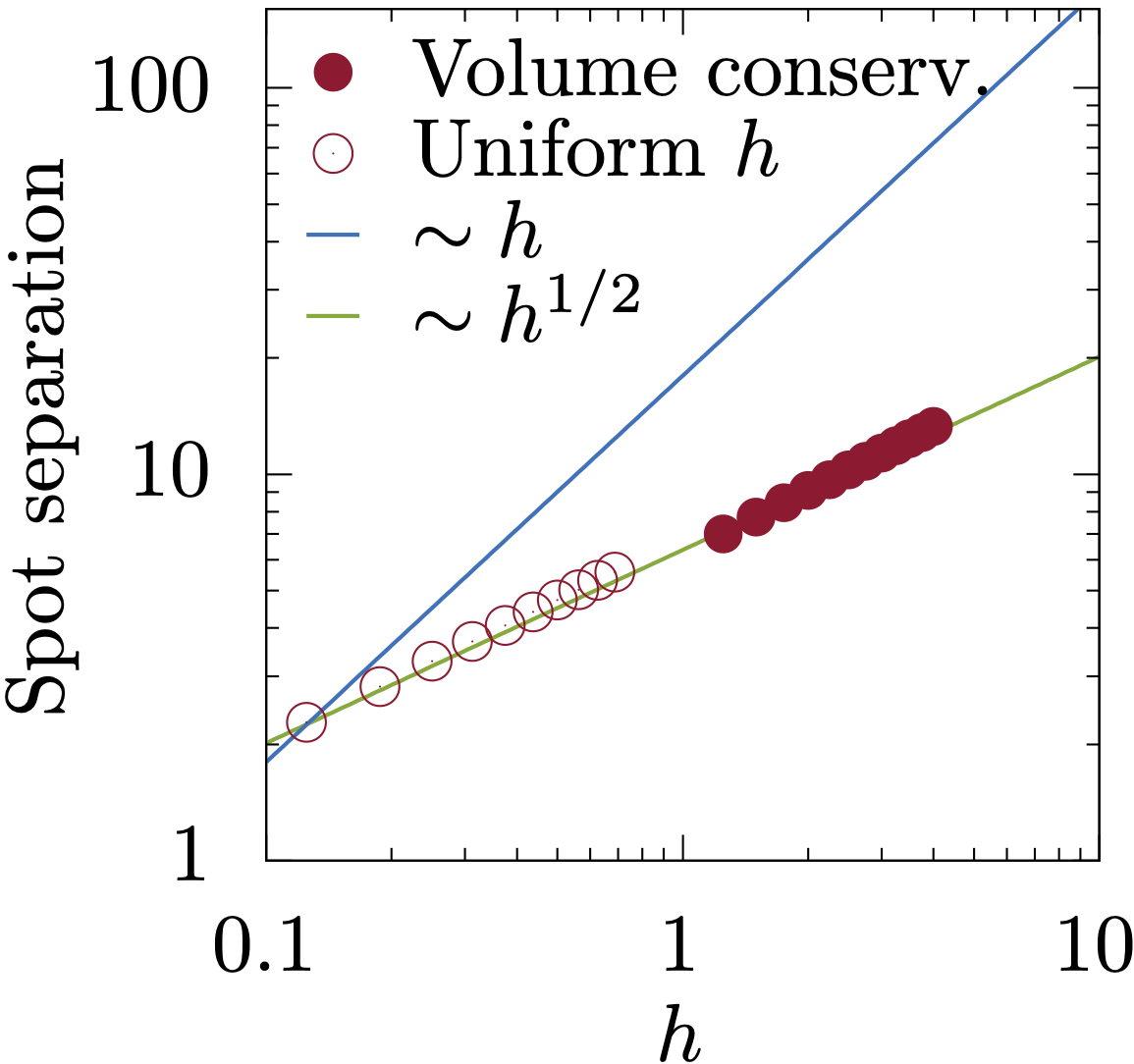}
		\end{tabular}
		\caption{Dependence of (a) spot size and (b) spot separation on the shell thickness, $h$, for the volume conserved ($h$ decreases with in-plane expansion) case and for uniform $h$ all over the shell at $D=0, c^\ast=0.1, R=30$. The plots are with respect to the initial undeformed value of $h$ in the volume conserved cases. }
		\label{fig:Spots}
	\end{figure}

	For this spotted pattern (that occurs, for example, at $c^\ast=0.1$), we estimate both the spot size and separation, and show how the pattern arising from the buckling instability scales with changing thickness. Assuming the shape of spots is close to circular, we calculate the average spot radius as $\sqrt{A_l/\pi N_{\textrm{sp}}}$, where $A_l$ is the total surface area of regions of the shell with low contractility, and $N_{\textrm{sp}}$ is the total number of spots. The results for simulated spheres of different thicknesses are depicted in Fig.\ref{fig:Spots}a along with linear and square root dependencies. The dependence of average spot radius on the initial shell thickness demonstrates the same  $\lambda \sim h^{1/2}$ scaling, both in the case of shells that maintain uniform thickness as well as those which conserve volume. The data is offset by a constant for the volume conservation cases, because the actual shell thickness is reduced at the spots. The pattern does not propagate when $h<0.125$ and $h<1.3$ for the uniform and nonuniform thickness, respectively, but the shells do exhibit the buckling instability at the location of initial perturbation. On the other hand, the shell is stable to small perturbations at large $h$ and returns to the initial spherical shape of uniform radius.
	
	We also estimate the distance between spots (in Fig.\ref{fig:Spots}b), which is computed as $2(\sqrt{A_{\textrm{tot}}/\pi N_{\textrm{sp}}}-\sqrt{A_{l}/\pi N_{\textrm{sp}}})$, where $ A_{\textrm{tot}}$ is the total shell area in deformed state. One can see that the spot separation in both cases is in a good agreement with \cite{Cerda2003} scaling as $\lambda \sim h^{1/2}$. Since the domain of high contractility remains at the initial thickness in both the cases considered, they show the same dependence on $h$.
	
	So far, the contractility change $\Lambda_{\textrm{max}}/\Lambda_{\textrm{min}}$ associated with biaxial expansion $\varepsilon$ was kept constant. However, $\varepsilon$ also determines the characteristic wavelength and the amplitude of wrinkles. Simulations at various $\Lambda_{\textrm{max}}/\Lambda_{\textrm{min}}$ show that a decreased ratio leads to a much larger spot size and separation, obeying the scaling $\lambda \sim \varepsilon^{-1/4}$, and thus, demonstrating the robustness of patterns to varying model parameters. We discuss the computational results in the Appendix \ref{appendix:c}. 
	
	We also note that the global structure of  the spotted pattern is constrained by topology in analogy with the packing and crystallization of particles on curved surfaces \cite{brojan2015wrinkling, lidmar2003virus, li2018large}. The wrinkling pattern can be considered as a triangular lattice tessellation of a spherical surface with bulges at the vertices, which allows only finite number of configurations with equi--spaced vertices as defined by the Euler characteristic \cite{coxeter1961introduction}. Any defect in this tessellation, for example a bulge with a different number of neighbors that makes the lattice irregular, causes an increase of elastic energy. Such a bulge changes its shape (the initial bulge in Fig.\ref{fig:SphereD0evo}a), or eventually merges with another bulge (small bulges at the secondary ring in Fig.\ref{fig:SphereD0evo}a), to reduce the elastic stress and accommodate the characteristic wavelength of spot separation. 
	On the other hand, we note that the ridge-like structures (Fig.\ref{fig:SphereD0evo}b) resemble scars on spherical crystals \cite{bausch2003grain}, but arise from a completely different physical mechanism. The scars are high-angle grain boundaries which develop as a row of alternating disclinations associated with positive and negative Gaussian curvature, whereas our linear rows of bulges do not require any crystalline order.
		
	We have thus shown that the pattern formation can be understood from the mechanics of elastic instabilities. Although, these are driven by local expansile stresses that are in turn induced by a propagating chemical concentration, considerable insight into the kinetic pathway and final patterns can be obtained by analyzing the buckling and wrinkling transitions in terms of thin shell elastic theory for a prescribed differential tension created by a fixed chemical gradient.
	
	\section{Discussion}
	
	Morphogenetic folding in biology encompass a wide variety of buckling mechanisms ranging from those driven by individual cell shape changes in local regions to those stemming from differential growth rates in neighboring tissue regions \cite{tozluoglu_19b}. While the details differ, all these processes require a spatial patterning of the tension or the growth rate in the tissue, which is typically accomplished by biochemical gradients and resulting gene expression in biology.
	
	It is conceivable that there are multiple biophysical pathways through which mechanics can affect the chemical profile that imprints mechanical gradients (and which we therefore term ``mechanogens''), such as through in-plane stresses that affect cell shape and membrane tension and therefore, the uptake and production of mechanogens \cite{dasbiswas_16}. Furthermore, a mechanogen can also in principle stimulate or suppress cell contractility, and thus lead to local contractions and expansions.  In the absence of direct experimental evidence of such a candidate mechanogen and its effect on cell mechanics in a specific biological setting, we explore, here one such plausible mechanism in a specific, simplified setting as proof-of-concept. 
	
	In our model system, wrinkling  results from differential changes in the in-plane area of the cells in response to mechanogens whose production is stimulated by tissue curvature. This positive mechanochemical feedback drives a propagating pattern of wrinkles on the surface of a thin elastic shell that are associated with both high curvature and high mechanogen concentration. We note that while this realizes a new pathway for obtaining a spotted pattern morphology on a spherical surface compared to those realized through purely mechanical means, such as by buckling of pressurized shells \cite{paulose_13, stoop_15}, or by purely chemical means, such as through phase separating chemical factors on the surface of shells modeling pollen grains \cite{lavrentovich_16}, the linear ridge-like morphology is very different.  Unlike ridges appearing in an elastic material that is attached to a compliant substrate and subjected to a large compression \cite{zang2012localized, cao2014harnessing}, we have shown that similar but high-aspect-ratio patterns can be developed in a single-layer unconstrained system due to dynamic localized expansion. Such patterns are of particular interest due to potential applications in tissue engineering and fabrication of functional surfaces, for instance hydrophobic coatings \cite{mammen2015functional}. This study also provides a possible pathway for diverse surface patterning in naturally occurring shells, for example the spikes, pores and ridges on pollen grains and viral capsids \cite{lavrentovich_16, Dharmavaram2017,lidmar2003virus}, and may also have implications for deformations of cell membranes which are fluid in-plane but are endowed with curvature elasticity  \cite{tamemoto2020pattern}. 
	
	Our mechanochemical model realizes an excitable medium, where a local initial stimulus is propagated through the medium because a localized chemical fluctuation results in a global mechanical response.  This is therefore a complementary proposal to that of Turing patterns which arise from the chemical interactions of a fast and a slow diffusing chemical species. We have shown that the mechanochemical feedback between \emph{one} chemical species which acts as a local activator, and the global mechanical response of elastic shell, can result in propagating and tunable pattern formation that does not require interactions between multiple chemical species with very different diffusivity.
	
	Given the complexity of biological tissue, we propose that a first experimental exploration of these ideas should be in an in vitro context, such as in organoids  (collections of cultured cells in vitro that mimic organs), where recent progress has been made in studying mechanical wrinkling \cite{karzbrun_18} or in cell cytoskeletal extracts such as actin gels embedded with contractile myosin motors \cite{ideses_18}, which can be combined with chemical gradients that induce differential mechanical stresses. The details of the mechanochemical feedback are likely to be different from that assumed here, but our modeling approach is general and could be easily modified to capture these scenarios.
	
	\section*{Acknowledgments}
	This work was supported by funding from the National Science Foundation: NSF-CREST: Center for Cellular and Biomolecular Machines (CCBM) at the University of California, Merced: NSF-HRD-1547848.
	We gratefully acknowledge computing time on the Multi-Environment Computer for Exploration and Discovery (MERCED) cluster at UC Merced, which was funded by National Science Foundation Grant No. ACI-1429783. We acknowledge insightful discussions with Stephanie H{\"o}hn, Nir Gov and Daniel Beller.
	
	\appendix
	\section{Details of discretized model} \label{appendix:a}
	
	The elastic energy defined by Eq.(\ref{Eq_ElastEnergy}) is discretized on a triangular mesh with node positions $\mathbf{X}_i$ and connecting bonds $l_{ij}$, and is written as 
	\begin{align}
		\mathcal{U}_{\textrm{d}} &= \frac{E}{2}\sum_\mathrm{nodes} V_i\Biggr( \frac{1}{2}\left(\frac{l_{ij} }{\overline{l}_{ij}}-1\right)^2 +  \left(\frac{h_{i} }{\overline{h}_{i}}-1\right)^2   + \frac{h_i^2 G_{i}^2}{12(1-\nu^2)} \Biggr),\label{Fe}
	\end{align}
	where $V_i={h}_i {A}_{i}$ is the node volume that assumed to be constant and calculated for the reference equilibrium configuration. The stretching energy is decomposed into in-plane and transversal strain, written in the first two terms, where the local thickness ${h}_{i}$ linearly depends on the observed area around a node ${A}_{i}=\sum A_j/3$, which is calculated as the average through adjacent triangles areas ${A}_j$. The first term accounts for deviations of the observed bond length $l_{ij}$ from the "optimal" length $\overline{l}_{ij}$ arising due to the elongation and shortening following local change in tissue contractility strength. This term is divided by 2 because each edge $l_{ij}$ is counted twice in the sum over all nodes. The second term accounts for thickness change and the optimal $\overline{h}_{i}$ determined by the condition of volume conservation $V_i=\overline{h}_i \overline{A}_{i}$, where $\overline{A}_{i}=\sum \overline{A}_j/3$. 

	\begin{figure}[b]
		\centering
		\includegraphics[width=0.31\textwidth]{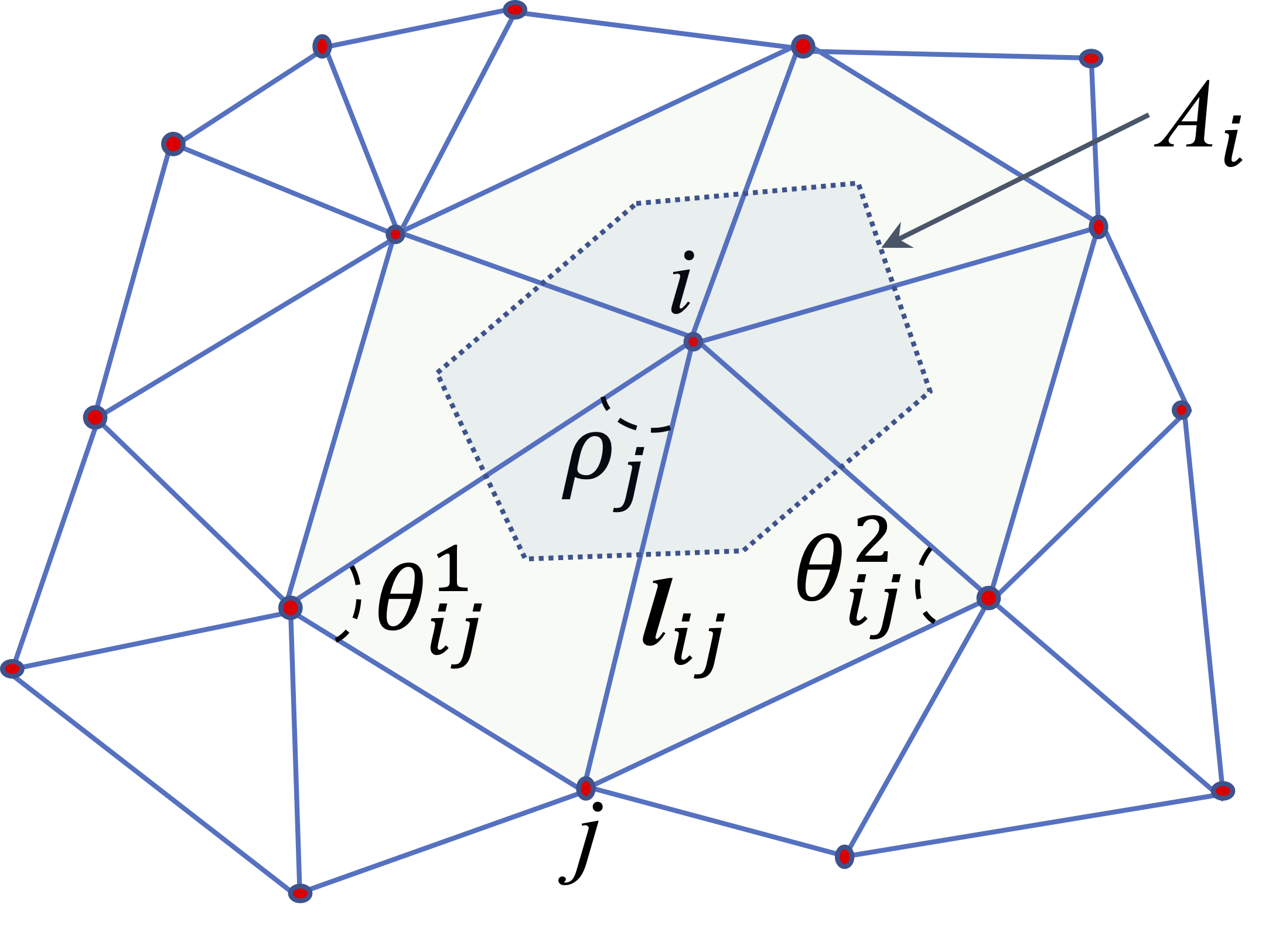}
		\caption{Geometry of discretized shell that is constituted as a collection of nodes with positions $\mathbf{x}_i$ connected by links ${l}_{ij}$, where each node spans an area $A_i$ around it.}
		\label{fig:mesh}
	\end{figure}

	The optimal lengths $\overline{l}_{ij}$ are associated with nonuniform in-plane contractility $\Lambda(c)$, which is prescribed by Eq.(\ref{Eq_Lambda}). For simplicity, we assume an inverse proportionality between contractility change and length shortening, and define this by  $\overline{l}_{ij}=l_{ij}^0(\Lambda^0/\Lambda(c))$,  where $l_{ij}^0$ is the reference length at the initial contractility $\Lambda^0$. 
	
	The last term in Eq.(\ref{Fe}) represents the bending energy, which vanishes in planar configuration if there is no natural curvature of the tissue $k_0=0$, and it is $k_0=1/R$ if cells develop wedged shapes forming a spherical shell of radius $R$ in equilibrium state. Following the Fopple-von-Karman approximation for the bending energy, the local curvature at each node is computed as $G_i^2= 4(H_i-k_0)^2-2(1-\nu)(K_i-2 H_i k_0+k_0^2)$, where the Gaussian curvature $K_i=(2\pi-\sum \rho_j)/A_i$ is expressed through the angles between two adjacent edges $\rho_j=\angle(l_{ij},l_{ij+1})$, and the mean curvature $H_i=||\sum ({l}_{ij}(\cot{\theta}^1_{ij}+\cot{\theta}^2_{ij}))||/(4 A_i)$ is defined over the adjacent edges, where $\theta^{k}_{ij}(k=1,2)$ are the two angles opposite to the edge in the two triangles sharing the edge $l_{ij}$ \cite{meyer2003discrete}. 
	
	Unlike transition to a wedge-like cell shape that generates a natural curvature of the shell \cite{hannezo_14}, we impose $k_0=1/R$ for a spherical shell of initial radius $R$, thus any deviation from the reference shape causes an increase in bending energy and can be generated by in-plane stress. 

	\begin{table}[t]
		\centering
		\begin{tabularx}{\columnwidth}{| b | X |}
			\hline
			Parameter & Description  \\
			\hline 
			$\nu=0.5$ & Poisson ratio \\
			$\Lambda_{\textrm{max}}=15$  & Maximal contractility\\ 
			$\Lambda_{\textrm{min}}=10$  & Minimal contractility\\
			$b=100$ & Width of transitional zone between domains with $\Lambda_{\textrm{max}}$ and $\Lambda_{\textrm{min}}$\\ 
			$H_s=0.05$ & Characteristic curvature at which the chemical production saturates \\
			$w=15$  & Production rate \\
			$\beta=1$  & Degradation rate \\
			\hline
		\end{tabularx}
		\caption{Parameters used in simulations}
		\label{table:1}
	\end{table}

	We assume overdamped dynamics and the positions of nodes $\mathbf{x}_i$ in the discretized shell are found by minimizing the elastic energy, $\mathcal{U}$, defined in the main text, by following the pseudo-time evolution equations:  $\gamma \partial \mathbf{x}_i/\partial t= - \delta \mathcal{U}_{\textrm{d}} / \delta \mathbf{x}_i$. Here, $\gamma$ is the friction coefficient associated with energy dissipation during elastic deformation, which can be caused by the surrounding fluid as well as viscous remodeling of the tissue material.
	
	\section{Numerical methods}\label{Section_NumM}
	
	We numerically investigate out-of-equilibrium dynamics of a deformed tissue sheet by solving Eqs.(\ref{Fe}) and (\ref{EqC}) using spatially unstructured finite-volume discretizations. The medium of uniform thickness in initial state is triangulated on a floating mesh consisting 25 000 nodes satisfying the Delaunay condition using the mesh generator Distmesh \cite{persson2004simple}. It allows to treat three-dimensional deformations following the physical displacement of the nodes sitting at vertices of triangles. The elastic energy (\ref{Fe}) is minimized over the positions of all nodes using the conjugate gradient algorithm with tolerance chosen as $10^{-4}\%$ of energy gain by executing a minimization step. The elastic energy gradients are calculated by sequential virtual displacement of nodes along each coordinate and then we perform iterative simultaneous update of all node positions. 
	
	The spatial derivatives in diffusion equation (\ref{EqC}) are calculated based on local approximation using the divergence theorem (Green-Gauss) gradient scheme \cite{ferziger2002computational}. We apply the node based method to compute concentration at nodes as the weighted average of all the neighboring triangles and then calculate gradients at the center of each triangle.  Then using Eq.(\ref{Eq_Lambda}) contractility at each node is computed that allows to find new optimal triangle side lengths respectively to given contractility. 
	
	\section{Patterning with large spots} \label{appendix:c}
	
\begin{figure}[t]
		\centering
		\includegraphics[width=0.475\textwidth]{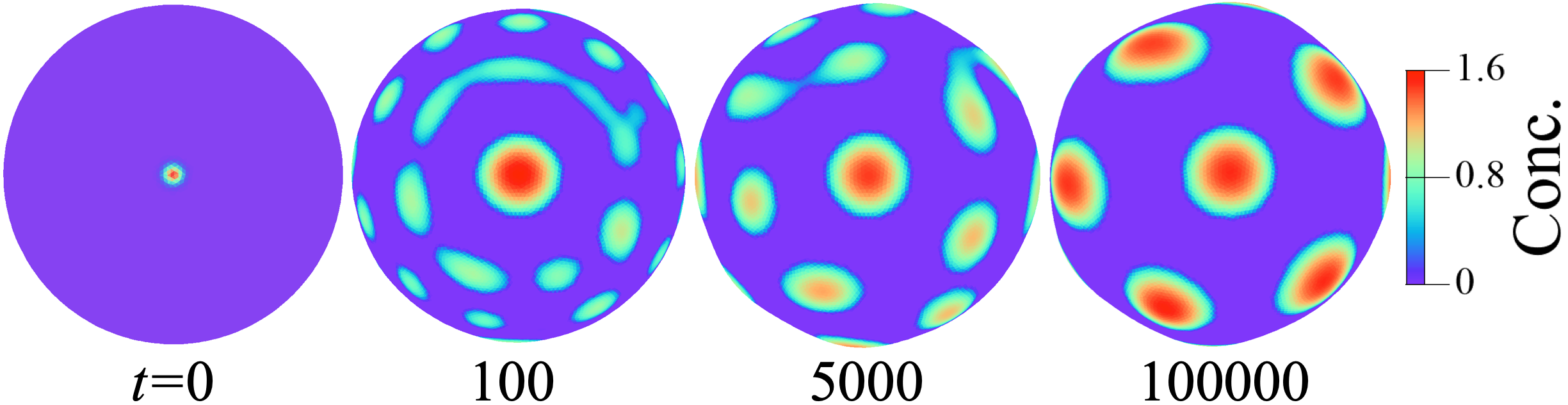}
		\caption{Simulation snapshots of spontaneous pattern formation on a spherical shell at $h=4, R=30, D=0$, decreased contractility change $\Lambda_{\textrm{max}}/\Lambda_{\textrm{min}}=1.05$ and concentration threshold $c^\ast=0.05$, but increased production rate $w=100$. The evolution reveals a fast phase of a spotted pattern formation followed by a slow coarsening. At the steady state ($t=100000$), only a few spots of high concentration with large radius remain, consistent with the theoretically predicted \cite{Cerda2003}, $\lambda \sim \varepsilon^{-1/4}$, scaling. }
		\label{fig:LargeSpots}
	\end{figure}

	In addition to the patterning on a spherical shell considered in the main part of this Communication, we explored the parameter space to demonstrate how the system can be tuned to produce target patterns and to study robustness of the mechanochemical feedback. Since the wavelength of wrinkling instability depends on both thickness and contractility change, the spot size and separation increase by reducing $\Lambda_{\textrm{max}}/\Lambda_{\textrm{min}}$ and increasing $h$. However, larger wavelength is associated with smaller amplitude of deformations, and so to make the pattern propagating, the threshold $c^\ast$ has to be lowered, and production rate $w$ increased to compensate for the decreased mean curvature in Eq.(\ref{EqC}). Starting with a single small point--like perturbation in concentration, the resulting patterns demonstrate only a few uniformly distributed large spots of circular shape (Fig.\ref{fig:LargeSpots}). Unlike patterning with smaller structures, here the time evolution reveals a fast phase when comparatively small patches of high concentration appear ($t=100$) and a subsequent slow coarsening that results in only several large, equi--spaced bulges on the shell surface. The global structure is constrained by topology and defined by the Euler characteristic \cite{coxeter1961introduction}. Thus, the sphere with large characteristic spot separation accommodates a finite number of spots at given sphere radius.  At $\Lambda_{\textrm{max}}/\Lambda_{\textrm{min}}=1.05, h=4$ and $R=30$, the spot separation is of the order of the sphere radius and the shell develops a regular icosahedral shape with 12 bulges at vertices.   Simulations at various $\Lambda_{\textrm{max}}/\Lambda_{\textrm{min}}$ demonstrate spot separation and radius that follow $\lambda \sim \varepsilon^{-1/4}$ scaling. For instance, at $\Lambda_{\textrm{max}}/\Lambda_{\textrm{min}}=1.15$ and $\Lambda_{\textrm{max}}/\Lambda_{\textrm{min}}=1.3$ the sphere demonstrates average spot separation $\sim 0.75 R$ and  $\sim 0.6 R$, respectively. Note that concentration saturation level is higher than in Fig.\ref{fig:SphereD0}-\ref{fig:unih} because of increased production $w$. 
	\\

	\FloatBarrier
	
	\bibliography{bibliography.bib}
	\bibliographystyle{ieeetr}
\end{document}